\documentclass[final,1p,times,authoryear]{elsarticle}
\usepackage{booktabs}
\usepackage[utf8]{inputenc}

\usepackage{amssymb}
\usepackage{lipsum}
\usepackage{float}
\usepackage{hyperref}
\usepackage{graphicx}	
\usepackage{amsmath}
\usepackage{soul}
\usepackage{color}

\usepackage{lineno}

\journal{Icarus}

\begin{document}

\begin{frontmatter}



\title{Estimates of Rotation Periods for Jupiter Trojans with the Zwicky Transient Facility Photometric Lightcurves}


\author[1]{Zhuofu (Chester) Li}
\author[1]{Yasin A. Chowdhury}
\author[1]{\v{Z}eljko Ivezi\'{c}}
\author[2]{Ashish Mahabal}
\author[1]{Ari Heinze}
\author[3]{Lynne Jones}
\author[4]{Mercedes S. Thompson}
\author[1]{Eric Bellm}
\author[1]{Mario Juri\'{c}}
\author[1]{Andrew J. Connolly}
\author[5]{Bryce Bolin}
\author[6]{Frank J. Masci}
\author[6]{Avery Wold}
\author[2]{Reed L. Riddle}
\author[2]{Richard G. Dekany}

\affiliation[1]{organization={Department of Astronomy and the DiRAC Institute, University of Washington}, 
                addressline={3910 15th Avenue NE}, 
                city={Seattle}, 
                postcode={98195}, 
                state={WA}, 
                country={USA}}
\affiliation[2]{organization={Caltech Optical Observatories, California Institute of Technology}, 
                city={Pasadena}, 
                state={CA}, 
                country={USA}}
\affiliation[3]{organization={Rubin Observatory}, 
                addressline={950 Cherry Ave.}, 
                city={Tucson}, 
                postcode={85719}, 
                state={AZ}, 
                country={USA}}
\affiliation[4]{organization={Department of Physics \& Astronomy, University of British Columbia}, 
                country={Canada}}
\affiliation[5]{organization={Eureka Scientific}, 
                addressline={2452 Delmer Street}, 
                city={Oakland}, 
                postcode={94602}, 
                state={CA}, 
                country={USA}; }
\affiliation[6]{organization={IPAC, California Institute of Technology}, 
                addressline={1200 E. California Blvd}, 
                city={Pasadena}, 
                postcode={91125}, 
                state={CA}, 
                country={USA}}

\begin{abstract}
We present new rotational period estimates for 216 Jupiter Trojans using photometric data from the Zwicky Transient Facility (ZTF), including 80 Trojans with previously unknown periods. Our analysis reveals rotation periods ranging from 4.6 hours to 447.8 hours. These results support the existence of a spin barrier for Trojans larger than 10 km, with periods clustering between 4 and 4.8 hours. This spin barrier is roughly twice as long as that observed for main-belt asteroids, suggesting that Jupiter Trojans have significantly lower bulk densities, likely due to a higher fraction of ices and volatile materials in their composition. We identify three new Trojans with reliable rotation periods near the spin barrier, doubling the number of known Trojans in this critical period range. Using these results, we estimate a mean density of $\sim 0.52~\mathrm{g/cm^3}$ for rubble-pile Trojans.
Our findings contribute to the growing body of evidence that many Trojans are rubble-pile bodies with distinct physical properties compared to main-belt asteroids. Looking forward, we anticipate that forthcoming data from the Vera C. Rubin Observatory’s Legacy Survey of Space and Time (LSST) will provide rotational period estimates for several hundred thousand Trojans, down to objects as small as 1 km, enabling a more detailed investigation of their rotational properties and internal structure.
\end{abstract}




\end{frontmatter}




\section{Introduction}
\label{introduction}

Jupiter Trojans, or Trojan asteroids, are a distinct population of small bodies that share Jupiter's orbit around the Sun, residing near the Lagrangian points L4 and L5. These points, leading and trailing Jupiter by approximately 60°, are two of five gravitational equilibrium zones created by the Sun–Jupiter system. At L4 and L5, small bodies like Trojans can remain in stable orbits for billions of years, forming what are known as the "leading" and "trailing" Trojan swarms. These regions have long been considered gravitational traps, preserving valuable clues about the early Solar System \citep{2002aste.book..725M, Bottke_2023}.

The origin of Jupiter Trojans has been debated extensively. Due to their sizes and distances from Earth, information on their formation and evolution remains limited. Trojans are thought to be more primitive than main-belt asteroids based on their compositions and physical properties \citep{2015aste.book..203E}. Earlier theories proposed that they formed near Jupiter as planetesimals and were captured during the planet’s formation \citep{2005Natur.435..462M, 2013ApJ...768...45N}, while more recent work suggests they may have originated farther out in the trans-Neptunian region and were scattered inward during the giant planet migration \citep{2013ApJ...768...45N}. This theory is supported by their low albedos and reddish colors, similar to Kuiper Belt objects \citep{2016AJ....152...90W, 2024PSJ.....5..193B}. Recognizing their scientific value, NASA's Lucy mission aims to explore a diverse selection of these primitive bodies through targeted flybys \citep{Levison_2021}.

Jupiter Trojans display a bimodal distribution in both color and surface characteristics—some are redder, others more neutral—which may reflect differences in their formation regions or evolutionary histories \citep{2011AJ....141...25E, 2014AJ....148..112W, 2016AJ....152...90W}. Their low bulk densities (around $10^3 \text{kg/m}^3$), comparable to cometary nuclei, suggest a composition of rock and ice \citep{2006Natur.439..565M, 2014ApJ...783L..37M, 2021PSJ.....2..170B, 2024SSRv..220...17M}. Additionally, the L4 swarm contains roughly $1.6\pm0.1$ times more objects than the L5 swarm \citep{2007MNRAS.377.1393S}, possibly due to differences in dynamical stability, collisional evolution, or migration processes \citep{2020MNRAS.495.4085H, 2023A&A...669A..68L}.

Understanding the rotational properties of Jupiter Trojans is critical to deciphering their physical structure and collisional history. A key feature is the "spin barrier"—the maximum rotation rate an object can sustain before centrifugal forces overcome self-gravity, potentially leading to structural failure or binary formation \citep{2007Icar..190..250P}. This limit is often explained by the cohesionless "rubble-pile" model, in which loosely bound aggregates rotate more slowly than solid, monolithic bodies \citep{Carbognani_2017}. For example, asteroids larger than ~0.15 km typically have rotation periods longer than 2.2 hours. Although C- and S-type asteroids have different bulk densities, only marginal differences in their spin barriers have been observed, suggesting that other factors may influence rotational stability.

Jupiter Trojans have a notably longer spin barrier—around 4 to 4.8 hours—about twice that of main-belt asteroids \citep{2014MSAIS..26...47M, 2015Icar..254....1F, 2017A&A...599A..44S, 2021PSJ.....2..191C}. This longer period points to lower bulk densities, likely due to higher fractions of ices and volatile materials \citep{2024SSRv..220...17M}. As of April 2024, approximately 240 reliable Trojan rotation periods have been reported by the Minor Planet Center \citep{Warner2009Icarus}, covering objects down to 50 km in diameter with 98\% completeness. However, this remains a relatively small sample. Expanding the dataset will help refine the spin barrier estimate and improve constraints on Trojan bulk density and internal composition—offering deeper insight into their origins and evolution.

To find new Trojan rotation periods, we utilize ZTF \citep{2019PASP..131a8002B}, a wide-field time-domain survey that offers high-cadence photometric data \citep{2019PASP..131g8001G}. ZTF operates on the Palomar P48 telescope, equipped with a 580-megapixel mosaic CCD camera and a 47 deg² field of view. It regularly scans the northern sky every ~2 days in SDSS-like $gri$ bands, reaching a 5$\sigma$ limiting magnitude of ~20.5 in the $r$-band \citep{2020PASP..132c8001D}. These features make ZTF well-suited for detecting Trojan lightcurves and estimating their rotation periods and bulk densities.

Looking to the future, we anticipate that our results will lay the groundwork for studies using data from the Vera C. Rubin Observatory’s Legacy Survey of Space and Time (LSST; \citealt{2019ApJ...873..111I}). LSST will transform our ability to study Jupiter Trojans and other small bodies in the Solar System, with plans to provide multi-band lightcurves for hundreds of thousands of Trojans, extending the size range down to objects as small as 1 km in diameter. The increased sample size and more detailed lightcurves from LSST will enable a comprehensive analysis of Jupiter Trojans' photometric and rotational properties, including color, size distribution, and rotational frequencies. This will allow us to better understand the diversity of their physical characteristics across different swarms and sizes. With LSST, we anticipate not only a significant increase in the number of known Trojans but also a deeper understanding of the processes that shaped the early Solar System and the forces that led to the formation and evolution of Jupiter Trojans.

In this paper, we explore the rotational characteristics of Jupiter Trojans using photometric data from the ZTF. In Section \ref{data_methods}, we describe our data acquisition, pre-processing, and lightcurve analysis procedures, including bias correction and rotation period estimation techniques. Section \ref{results} presents our findings on the distribution of Trojan rotation periods and their connection to the spin barrier, offering insights into the structural and compositional characteristics of these objects. In Section \ref{conclusion}, we place our results in the context of current Trojan research and discuss how future data from the Vera C. Rubin Observatory's Legacy Survey of Space and Time (LSST) may enhance our understanding of their rotational dynamics. We also highlight future directions for investigating the formation, evolution, and internal structure of Jupiter Trojans.

\begin{figure}[t]
\centering
\includegraphics[width=0.6\linewidth]{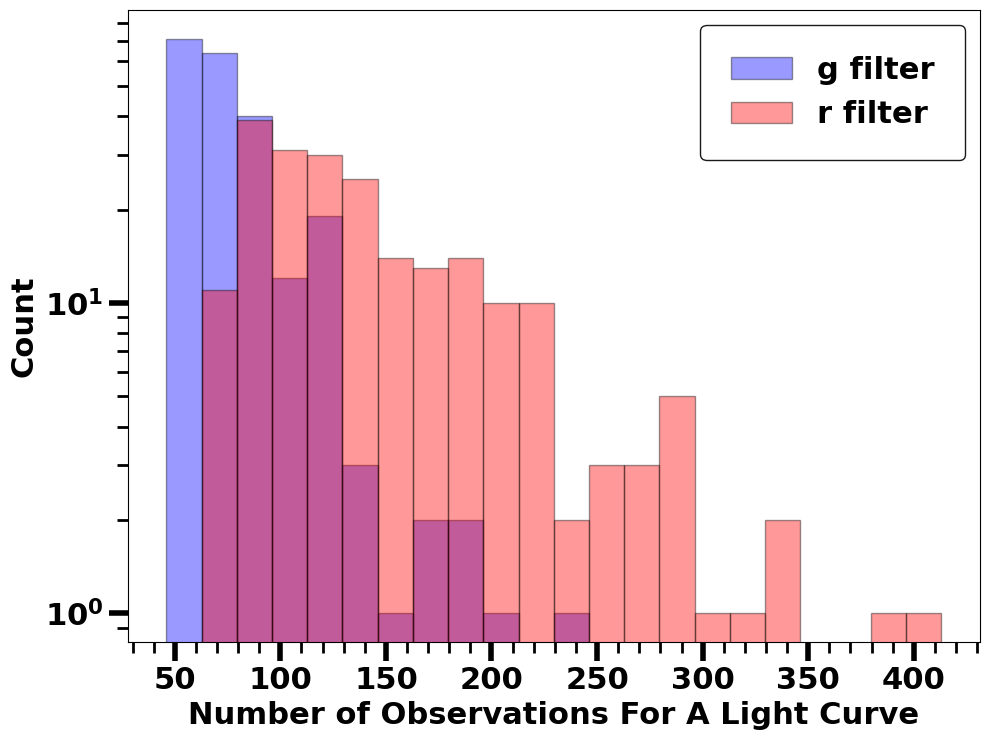}
\caption{Histogram showing the distribution of ZTF photometric observations for 2,813 Jupiter Trojans in the g-band (blue) and r-band (red). The x-axis represents the number of observations per object, while the y-axis (logarithmic scale) shows the number of objects with that count. Most objects have fewer than 100 observations in either band, with a steep drop-off at higher counts. A small fraction of objects have over 200 observations, particularly in the r-band. The darker pink color in the histogram indicates an overlap between the g- and r-band distributions. The difference in observation counts reflects the greater number of r-band exposures, consistent with ZTF’s observing strategy.}
\label{fig:histogram_full_sample}
\end{figure}

\begin{figure*}[t]
\centering
\includegraphics[width=0.6\linewidth]{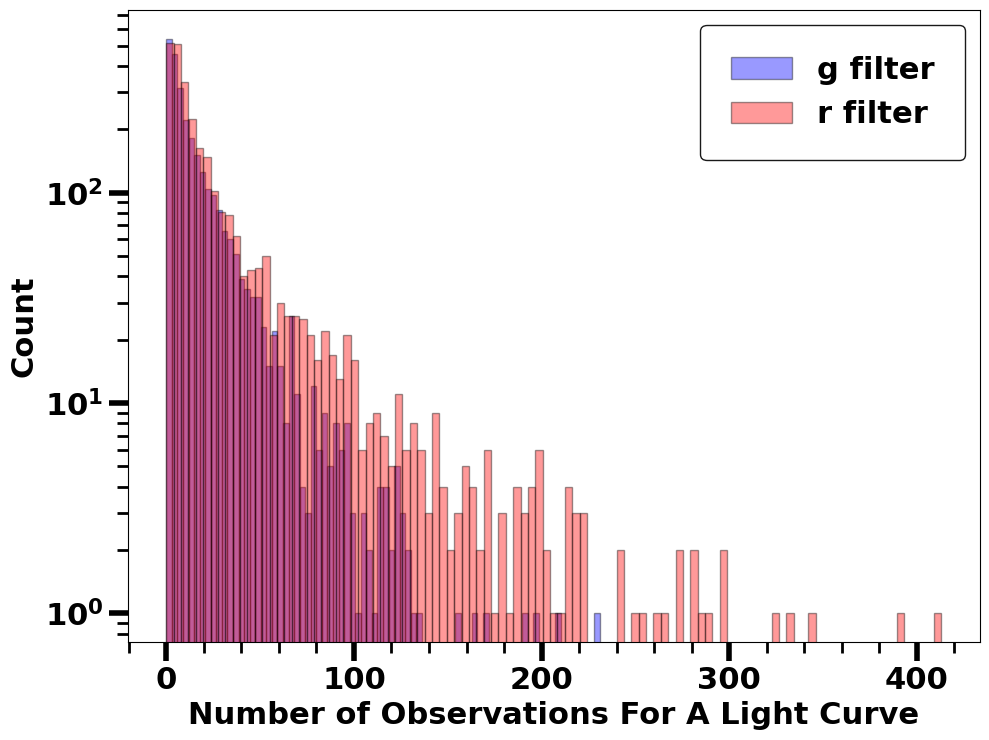}
\caption{Histogram showing the distribution of ZTF photometric observations for the final sample of 216 Jupiter Trojans in the g-band (blue) and r-band (red). The x-axis represents the number of observations recorded for each object, while the y-axis (on a logarithmic scale) indicates the number of objects with that observation count. This subset includes only objects with at least 130 total observations across both bands, and a minimum of 45 observations in each band. The darker pink color in the histogram represents overlapping bins between the g- and r-band distributions, providing a clearer view of the well-observed Trojans in the sample.}
\label{fig:histogram_final_sample}
\end{figure*}

\begin{figure*}[t]
\centering
\includegraphics[width=0.9\textwidth]{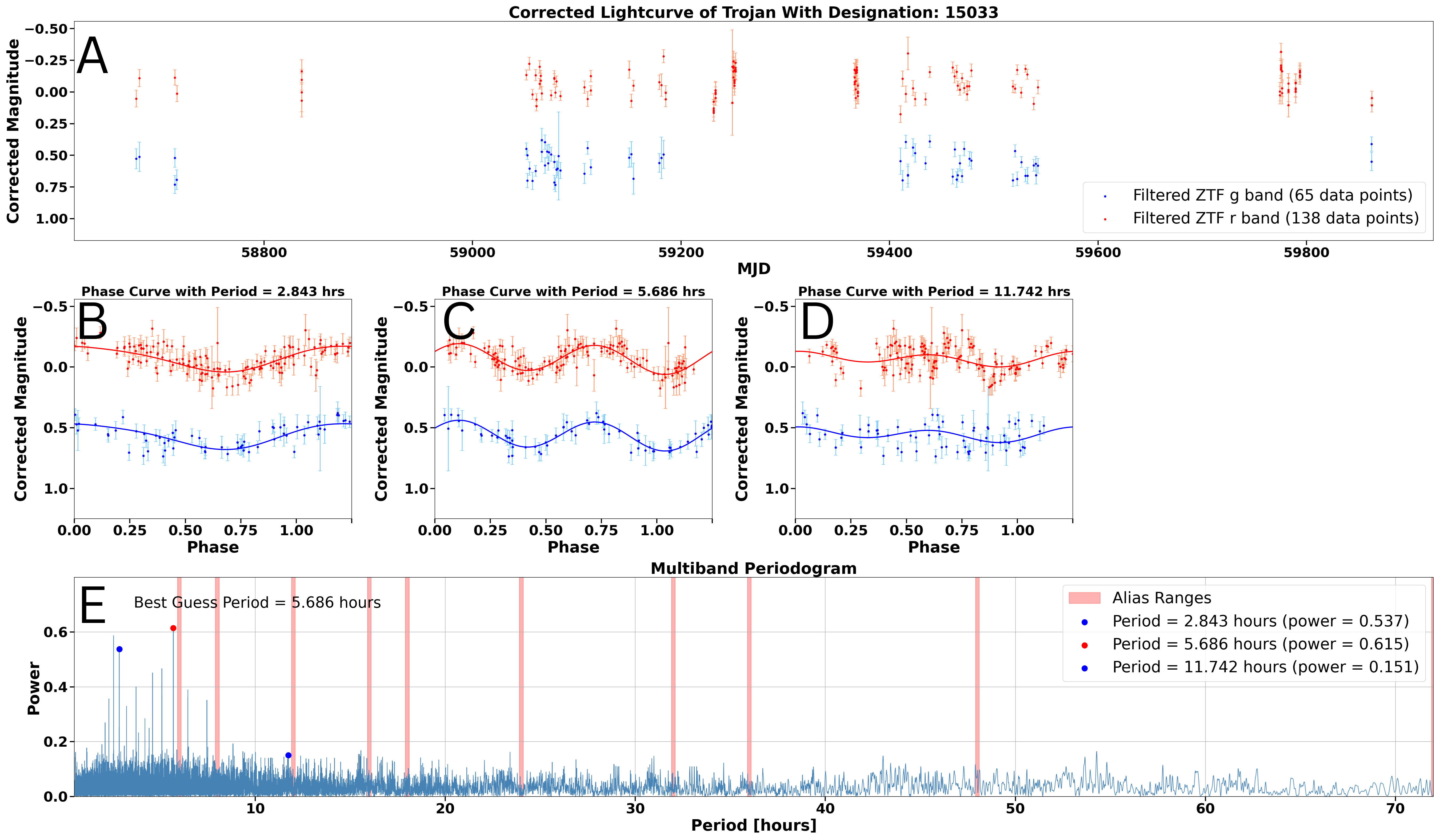}
\caption{Lightcurve Analysis for Trojan 15033. This is a 5-panel figure: panel A shows the corrected lightcurve for the ZTF g-band (blue) and r-band (red) data, plotted against Modified Julian Date (MJD). The lightcurve reveals variability across multiple observation epochs. Panels B, C, and D present phase-folded lightcurves for three potential periods derived from the Lomb-Scargle periodogram: (B) 2.843 hours, (C) 5.686 hours, and (D) 11.742 hours. In each case, the data is folded over the corresponding period, showing two minima, typical of rotating objects. The filtered g-band and r-band data are plotted with the number of minima indicated in the legend. Panel E shows the multiband periodogram, where the best-fit period of 5.686 hours is highlighted as the most likely rotational period. Alias ranges are shown as discussed in \ref{sec:data_analysis}.}
\label{fig:15033}
\end{figure*}

\section{Data and Methods}
\label{data_methods}

In this section, we provide a detailed overview of the data used and the methods employed to estimate the rotation periods of Jupiter Trojans from photometric lightcurves obtained by the ZTF. Specifically, we explain the steps taken to acquire the lightcurves, pre-process the data to remove biases and ensure quality, and the techniques applied to derive reliable rotation periods.

\subsection{Data Acquisition}

For this study, we utilized photometric data from the ZTF for Trojan lightcurves analysis. Specifically, we used data from the ZTF $r$-band and $g$-band filters, while excluding the $i$-band data. The decision to exclude the $i$-band was due to its significantly lower cadence and the lower quality of observations in this band, which were insufficient for accurately determining rotation periods. The $g$ and $r$ photometric bands, on the other hand, provided reliable time-series data with the necessary temporal coverage and photometric precision to support our analysis.

The ZTF's observational strategy involves nightly scans of the sky, complemented by morning and evening twilight observations that prioritize Solar System objects. ZTF's automated system identifies asteroids, including Jupiter Trojans, as part of its regular operations. 
The observation frequency for a given Jupiter Trojan in the ZTF dataset varies depending on the object's position, visibility, and observing conditions. In Section \ref{sec:data_process},  we discuss the method for correcting the lightcurve variability due to the viewing geometry. 
Observations of these objects are stored in a comprehensive depot that includes data from all three asteroid-specific observation periods: regular nightly observations, morning runs, and evening runs. This depot, which contains data on all known asteroids observed by ZTF, served as the primary source for our study.

To extract relevant lightcurves for Jupiter Trojans, we cross-matched a list of known Trojans, compiled in October 2022, with the ZTF asteroid depot. The list of Trojans was drawn from the MPC database, which tracks the orbits of all known small bodies in the Solar System \citep{Williams2015}. Through this cross-matching process, we identified 2,813 Jupiter Trojans with sufficient observational coverage in the $g$ and $r$ bands. The lightcurves of these objects formed the basis of our analysis.

\begin{figure*}[t]
\centering
\includegraphics[width=0.9\textwidth]{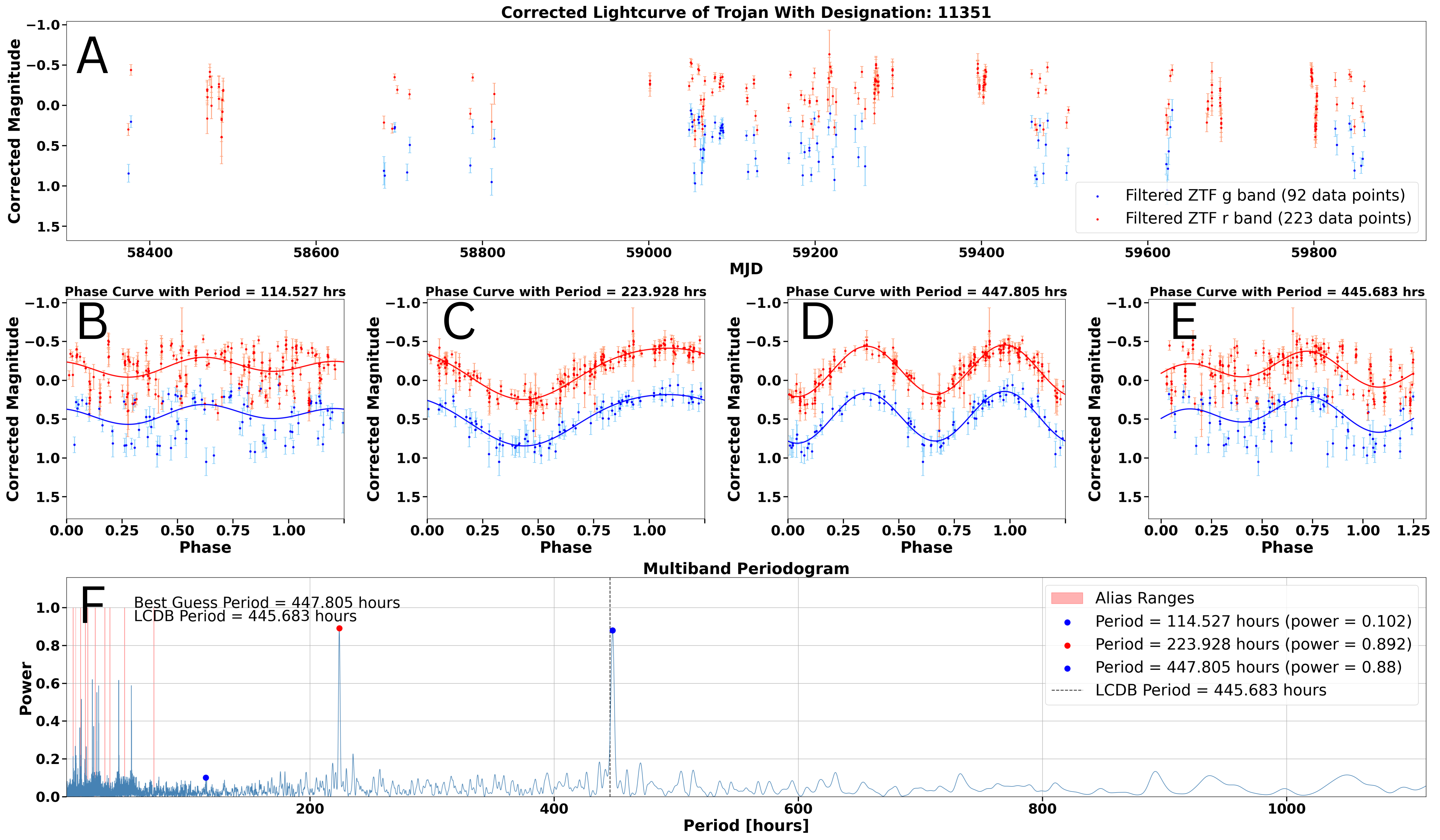}
\caption{Lightcurve Analysis for Trojan 11351 (Leucus). This is a 6-panel figure: panel A shows the corrected lightcurve for the ZTF g-band (blue) and r-band (red) data, plotted against Modified Julian Date (MJD). This lightcurve displays variability over time, with clear differences between the two bands. Panels B, C, and D present phase-folded lightcurves for three potential periods derived from the Lomb-Scargle periodogram: (B) 114.527 hours, (C) 223.928 hours, and (D) 447.805 hours. Panel E shows the phase-folded lightcurve for the LCDB (Lightcurve Database) period of 445.683 hours. The phase-folding of the data provides insight into the periodic nature of the Trojan’s rotation, with panels B and D showing two minima, characteristic of a rotating body, while panel C exhibits a single minimum. Panel F displays the multiband periodogram, where the best-fit period of 447.805 hours is indicated as the most likely period. This period is very close to the LCDB period of 445.683 hours, suggesting consistency with prior observations. Alias ranges are shown as discussed in \ref{sec:data_analysis}.}
\label{fig:11351}
\end{figure*}

\begin{figure}[t]
\centering
\includegraphics[width=0.49\textwidth]{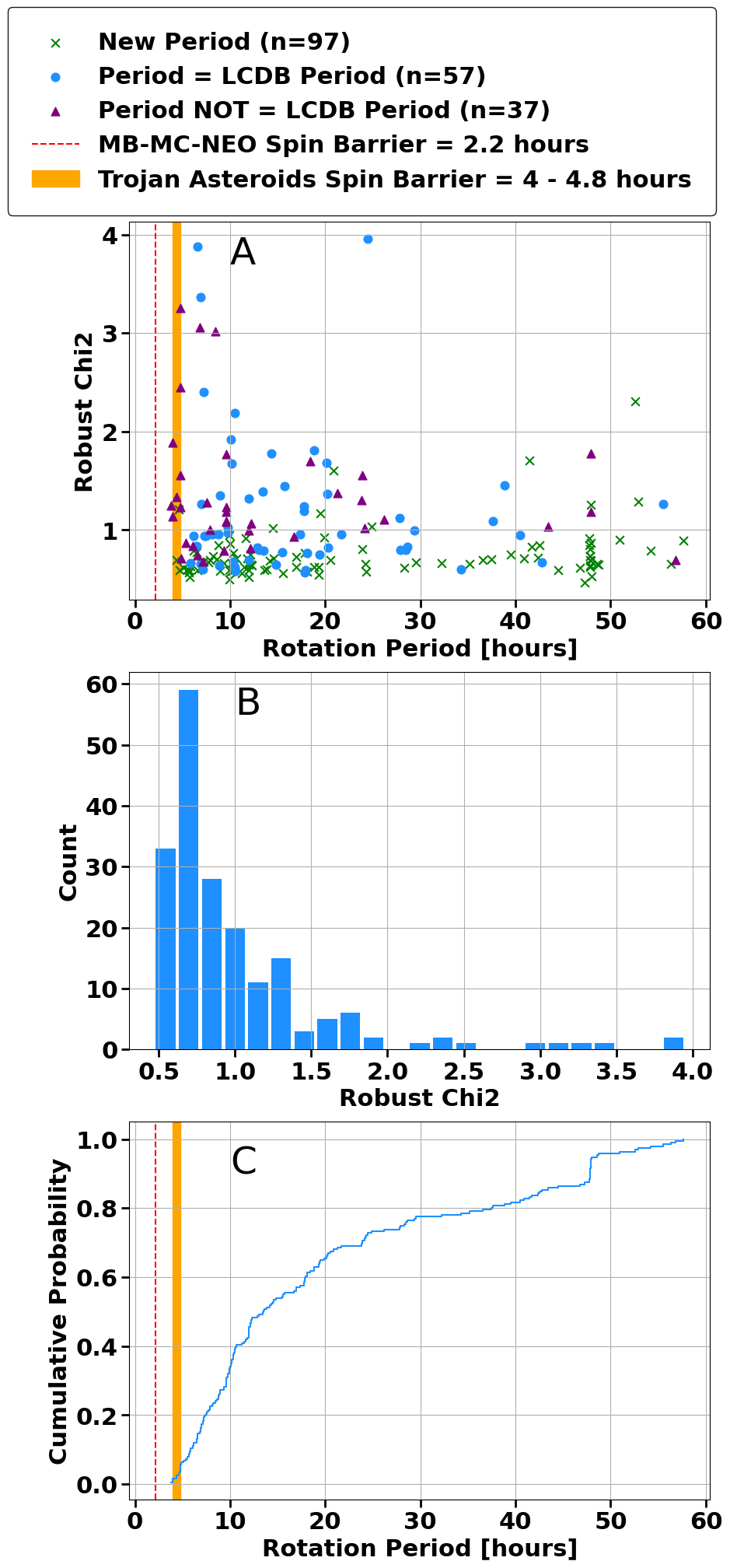}
\caption{
Panel A shows the Chi-square of the computed periods versus the computed periods for the 191 selected Jupiter Trojans with rotation periods less than 60 hours. The data points are grouped into three categories based on their period comparison to previously known periods. Lower Chi-square values indicate a better fit between the model and observed data. Panel B presents a histogram of the Chi-square values for the entire sample, showing the distribution of how well the computed periods fit the observed data. A peak at low Chi-square values suggests that most of the computed periods provide a good fit, although a tail toward higher values indicates some objects with less reliable period determinations. Panel C shows the cumulative distribution of computed periods. The dashed red vertical line and solid orange line mark two spin barrier limits: one for Main Belt Asteroids, Mars-Crossers, and Near-Earth Objects (MB-MC-NEO) at around 2.2 hours, and the other for Jupiter Trojans at around 4 to 4.8 hours.}
\label{fig:summary}
\end{figure}

For each identified object, ZTF provided calibrated photometric measurements using zero-point and color-term adjustments based on the Pan-STARRS1 photometric catalog \citep{Pan_starrs1}.
The photometry from ZTF is described in detail by \cite{2019PASP..131a8002B} and \cite{2019PASP..131a8003M}. These lightcurves represent the variation in brightness of each Trojan over time and are essential for determining their rotation periods. In the subsequent sections, we describe the pre-processing steps taken to ensure data quality and the methodology used to derive reliable rotation period estimates from these lightcurves.

\subsection{Lightcurve Pre-Processing}
\label{sec:data_process}

Before analyzing the data, we performed several pre-processing steps to ensure that the lightcurves accurately reflected photometric variability caused by asteroidal rotation, while minimizing the impact of external factors such as observational biases or variations in object distance.

As seen in Figure~\ref{fig:histogram_full_sample}, which shows the distribution of ZTF observations for our initial sample of 2,813 Jupiter Trojans, there is a clear variation in the number of observations per object. The x-axis represents the number of observations for each object, and the y-axis (logarithmic scale) shows the count of objects with a given number of observations. Most objects have fewer than 100 observations in either the $g$ or $r$ bands, with a sharp decline beyond 100. We also observe that the $r$-band generally has more observations per object compared to the $g$-band, a common feature in ZTF data due to its observing strategy, which tends to prioritize $r$-band exposures. This variation in observation count indicates that a significant portion of our sample may not have sufficient data points for reliable rotational analysis.

To address this, we applied a filtering criterion to select objects with a minimum of 130 total observations across both the $g$-band and $r$-band, with at least 45 observations in each band. This threshold was chosen to ensure adequate temporal coverage for detecting rotational variations and computing reliable rotation periods. 
This criterion does not depend on whether the observations were from the same or multiple oppositions since our lightcurve correction method is robust across opposition epochs.
The selection criterion is based on the distribution of $g$-band observations, as there is a noticeable cut-off at around 130 observations, beyond which the number of objects sharply decreases. By applying this filter, we reduced our dataset to 216 Trojans.

Figure~\ref{fig:histogram_final_sample} illustrates the distribution of ZTF observations for the final sample of 216 Trojans that met the observation threshold. Similar to Figure~\ref{fig:histogram_full_sample}, the x-axis shows the number of observations, and the y-axis (logarithmic scale) indicates the count of objects. The median number of observations in the $g$-band is 69, while in the $r$-band it is 125. This subset of objects represents those with the most well-sampled lightcurves, providing sufficient data points to accurately estimate rotation periods.

In the next step of our analysis, we corrected the observed magnitudes for heliocentric distance, geocentric distance, and phase angle.This correction was crucial for isolating the intrinsic rotational lightcurves, ensuring that the brightness variations were caused by the rotation of the asteroids, rather than changes in their distance from Earth or the Sun. To perform the correction, we retrieved the predicted $V$-band magnitudes for each object from the Minor Planet Center (MPC, \cite{Warner2009Icarus}) based on their expected brightness at the specific observation times. The MPC's predicted $V$-band magnitudes have already incorporated the effects of heliocentric distance, geocentric distance, and phase angle.
MPC calculates predicted V-band magnitudes for minor planets using the H-G magnitude system \citep{1989bowell}. However, this system often falls short in accurately modeling the phase curves of Trojan asteroids, which exhibit a minimal opposition effect \citep{2010Trojan_Oppo}. The more advanced H, $G_1$, $G_2$ photometric system developed by \citep{2010Muinonen} offers a better representation of asteroid phase curves. Consequently, relying on MPC magnitude predictions may lead to inaccuracies in capturing the phase behavior of asteroids with unique characteristics.

The correction process involved subtracting the predicted $V$-band magnitudes from the observed magnitudes at each time point, followed by a normalization step. In this normalization step, we subtracted the median of the resulting differences to ensure that the lightcurves reflected the asteroidal rotation, free from phase angle or distance-induced effects. This method allows us to analyze the variability in the lightcurves due to the rotation of the Trojans.


\begin{table*}
    \centering
    \begin{tabular}{ccccc}
        \toprule
        & \textbf{New Period} & \textbf{LCDB Period} \\
        \midrule
        \textbf{Reliability 1} & 27 & 15 \\
        \textbf{Reliability 2} & 27 & 24 \\
        \textbf{Reliability 3} & 58 & 65 \\
        \midrule
        \textbf{Total} & 112 & 104 \\
        \bottomrule
    \end{tabular}
    \caption{Reliability Flags Summary Table for the 216 Jupiter Trojans analyzed. The table shows the total number of reliability flags assigned to two groups of Trojans: those with known LCDB periods and those with newly computed periods from this study. Reliability flags are categorized into three levels: Reliability 3 indicates the highest confidence in the period determination, where the lightcurves show clear and well-defined minima, and the periodogram has a sharp peak. Reliability 2 represents periods that are presumed reliable but may require further investigation due to potential uncertainties in the data or periodogram results. Reliability 1 marks periods that are unreliable and need more data or more careful re-analysis. For each group, the table provides the number of objects in each reliability category. The total number of observations (N obs) and the amplitude of the lightcurve variation, along with the diameter of each object, are also included for further context. This categorization helps identify which periods are most reliable for further analysis and comparison with prior studies.}
    \label{tab:reliability}
\end{table*}

\subsection{Rotation Period Determination}

\label{sec:data_analysis}

Before searching for periodic signals in the ZTF lightcurves, we excluded data points with large magnitude errors (greater than 0.6 magnitudes), which could introduce noise or inaccuracies into the period analysis. This filtering step ensured that only high-quality photometric measurements were included in the subsequent analysis.

We then utilized the Multi-Band Lomb-Scargle periodogram \citep{2018ApJS..236...16V} to search for periodic signals by combining $g$-band and $r$-band ZTF observations. The Lomb-Scargle periodogram is a widely-used tool in time-domain astronomy for identifying periodic signals in unevenly spaced data, making it ideal for our dataset, which includes observations from two filters taken at irregular intervals. By analyzing both $g$ and $r$ bands simultaneously, we increased the sensitivity of our period detection, improving the likelihood of identifying true rotational periods.

After running the Multi-Band Lomb-Scargle periodogram, we removed specific ranges of periods that are known to be susceptible to aliasing effects due to the observational cadence of ZTF \citep{2021MNRAS.505.2954C}. These alias ranges, shown in Figure~\ref{fig:15033} panel E, include periods around 5.9 to 6.1 hours, 7.9 to 8.1 hours, and ranges between 71.9 to 72.1 hours. Removing these ranges helped reduce false-positive detections caused by regular sampling intervals in the ZTF data, which could mimic true periodic signals.

Following this step, we identified the periodic signal with the highest power in the Lomb-Scargle periodogram. Panel E in Figure~\ref{fig:15033} shows the multiband periodogram for Trojan 15033, where the best-fit period of 5.686 hours is highlighted as the most likely rotational period. The periodogram provides insight into the periodic nature of the lightcurve, with higher power values indicating stronger confidence in the detected period. Alias periods at 2.843 hours and 11.742 hours are also shown but have lower power, indicating they are less likely to be the true period.

However, the detected period may not necessarily represent the true rotation period. To ensure that the periodic signal corresponds to the rotation of the asteroid, we conducted further analysis by examining the halved and doubled periods of the best-fit signal. Panels B, C, and D of Figure~\ref{fig:15033} present the phase-folded lightcurves for the three candidate periods: 2.843 hours, 5.686 hours, and 11.742 hours, respectively. In each case, the lightcurve is folded over the corresponding period, and we plot the filtered $g$-band and $r$-band data, with error bars representing the photometric uncertainties.

A key feature of rotational lightcurves for ellipsoidal asteroids is the presence of exactly two minima per rotation, while more complex or irregular shapes may produce asymmetric or multi-peaked lightcurve profiles. To assess which of the candidate periods best fits the expected profile of an asteroid's rotational lightcurve, we applied a robust variant of the \(\chi^2\) test to evaluate the goodness of fit between the observed magnitudes and the modeled lightcurves. This adaptation used the interquartile range of the residuals to estimate the standard deviation, making the test less sensitive to outliers compared to the traditional \(\chi^2\) test. The lowest \(\chi^2\) value indicates the best match between the model and the observed data.

In addition to the \(\chi^2\) test, we evaluated the number of minima in the phase-folded lightcurves. A lightcurve that exhibits a double-peaked profile with two minima is consistent with the expected rotational signature of an asteroid. In Panels B, C, and D, the number of minima is indicated in each plot, and we observe that the lightcurve corresponding to a period of 5.686 hours (Panel C) has two clear minima, while the others show either a single minimum or less distinct features.

Using the combination of the robust \(\chi^2\) values and the visual inspection of the number of minima, we determined the most likely true rotation period for each Trojan. In this example, the period of 5.686 hours was selected as the true rotation period, as it resulted in a lightcurve with exactly two minima and the lowest robust \(\chi^2\) value.

To validate our methodology, we compared the determined rotation periods to previously known periods reported in LCDB\citep{Warner2009Icarus}. This cross-check provided additional confidence in the accuracy of our results.

Finally, we manually verified the results for the final sample of 216 Trojans. For each object, we confirmed the presence of exactly two minima in the phase-folded lightcurve and examined the periodogram for a clear peak at the derived period. A sharp peak in the periodogram, such as the one shown in Figure~\ref{fig:15033} panel E, indicates a strong periodic signal, increasing our confidence in the estimated period. For each object, we assigned a reliability flag, as detailed in Table~\ref{tab:reliability}, with flags of 3 indicating highly reliable periods, 2 for presumptively reliable periods, and 1 for unreliable periods requiring further investigation.


\begin{figure*}[t]
\centering
\includegraphics[height=0.39\textheight]{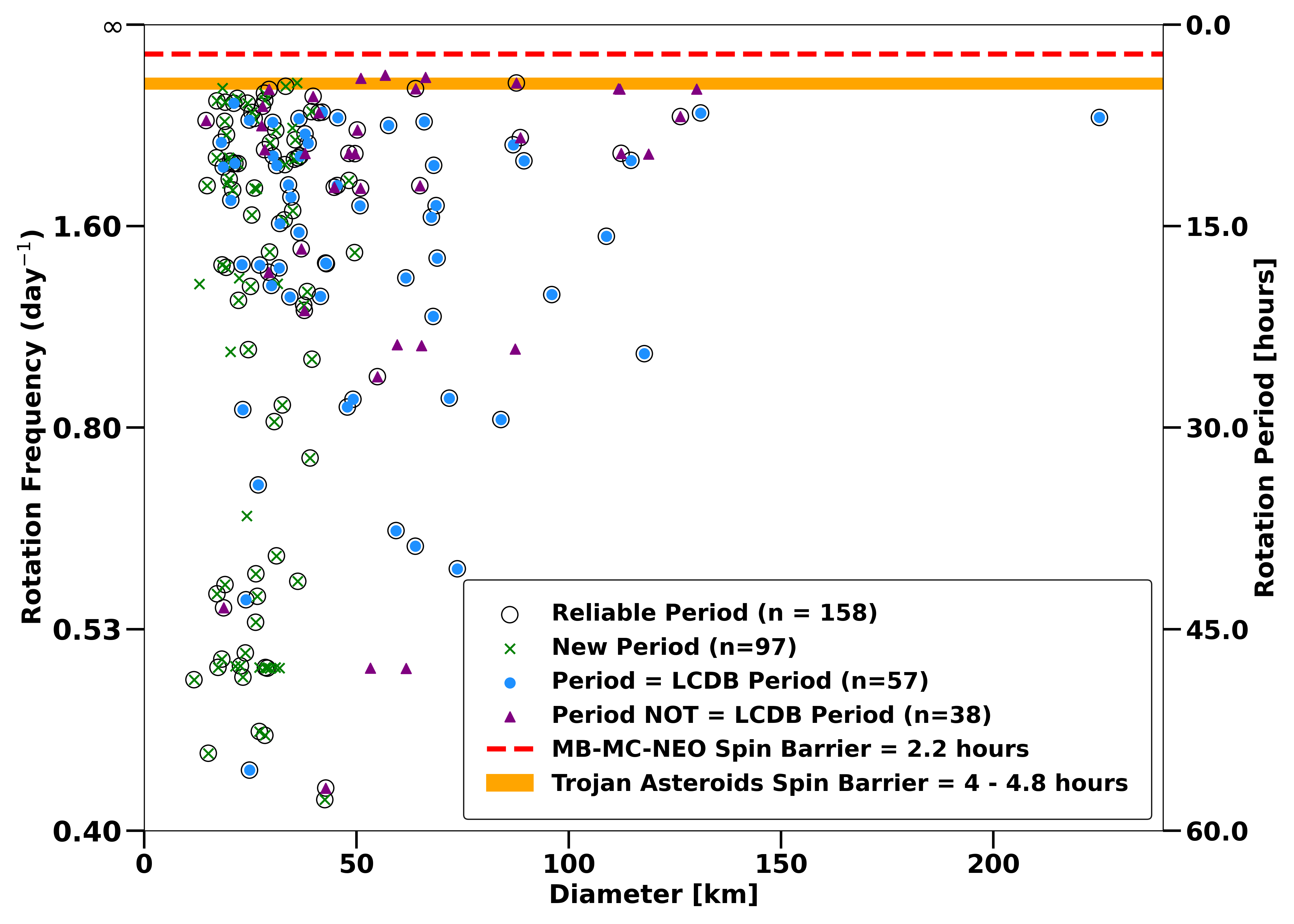}
\caption{This figure presents the relationship between the computed rotational frequency/periods of Jupiter Trojans and their diameters. Panel A shows the rotational frequency (y-axis) plotted against the diameter (x-axis) for the sample of 216 Jupiter Trojans. The data points are categorized into three groups: Trojans without previously known periods (green crosses), Trojans with computed periods that match within 10\% of their previously known periods (blue circles), and Trojans with computed periods that differ significantly from their previously known periods (purple triangles). The legend clearly distinguishes these groups to allow for easy interpretation of the reliability and novelty of the computed periods. The plot shows two spin barriers: one for Main Belt Asteroids, Mars-Crossers, and Near-Earth Objects (MB-MC-NEO) at around 2.2 hours, and another for Jupiter Trojans at approximately 4 to 4.8 hours. These barriers, represented by a dashed red line and a solid orange line, indicate the maximum rotational frequency for objects of different populations, with the Trojans' spin barrier shifted to lower rotational frequencies (longer periods). This suggests that Trojans have lower bulk densities compared to the MB-MC-NEO population. The objects plotted close to the spin barrier are of particular interest, as they provide insights into the structural integrity of these bodies. The plot shows that Trojans with smaller diameters generally exhibit shorter rotational periods (higher frequencies), but few objects have periods close to the spin barrier, reinforcing the idea of a rubble-pile structure for many Trojans, which prevents them from spinning faster without breaking apart. The absence of smaller Trojans with fast rotations further supports this structural model. Trojans without diameter measurements are not shown in this plot.}
\label{fig:spin_barrier}
\end{figure*}

\begin{figure*}[t]
\centering
\includegraphics[height=0.39\textheight]{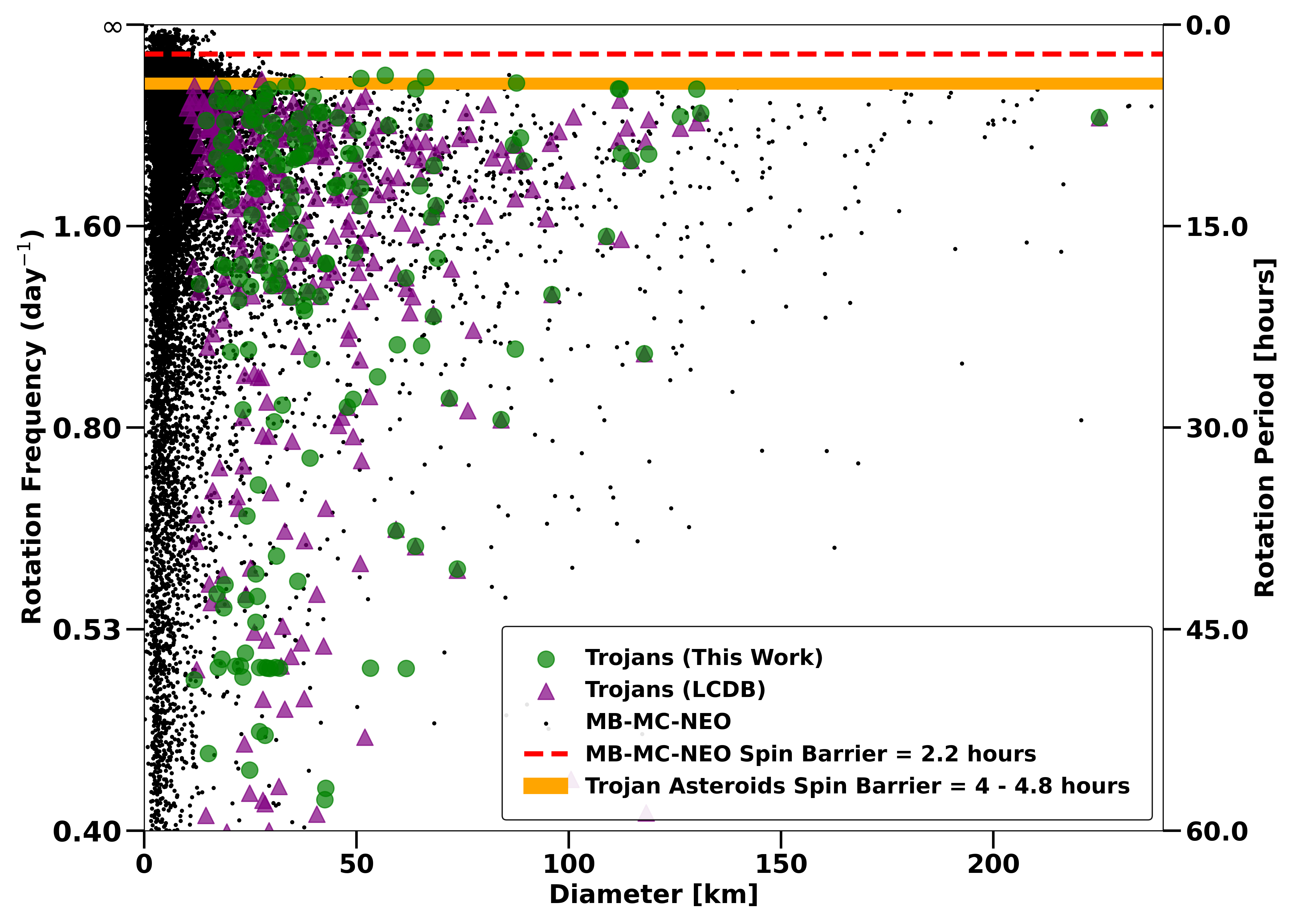}
\caption{This figure compares the rotational frequency/periods of Jupiter Trojans analyzed in this study with those of Trojans, Main Belt Asteroids, Mars-Crossers, and Near-Earth Objects (MB-MC-NEO) from the Asteroid Lightcurve Database (LCDB; \cite{Warner2009Icarus}). The y-axis shows the rotational frequency (inverse of the period), and the x-axis represents the diameter of the objects in kilometers. The data points are categorized into three groups: Trojans with newly computed periods from this study (green circles), Trojans with computed periods within 10\% of their previously known LCDB periods (purple triangles), and MB-MC-NEO objects (magenta squares). The plot also shows two spin barriers: the first for MB-MC-NEO objects at around 2.2 hours, and the second for Jupiter Trojans at 4 to 4.8 hours, both represented by dashed lines. The spin barrier for Jupiter Trojans is shifted to lower rotational frequencies (longer periods), which suggests that Trojans generally have lower bulk densities compared to MB-MC-NEO objects. The objects that plot close to the spin barriers, particularly those with new period estimates, are of significant interest because they help refine our understanding of the physical characteristics of these populations. From the plot, it is evident that smaller objects (those with diameters below 50 km) tend to have higher rotational frequencies, while larger objects rotate more slowly. The presence of Jupiter Trojans with long periods near the spin barrier provides strong evidence for the rubble-pile structure, as these bodies are prevented from rotating faster without breaking apart due to their low cohesive strength. MB-MC-NEO objects show a distinctly different distribution, confirming the differences in physical properties between these two populations.}
\label{fig:spin_barrier_all}
\end{figure*}

\begin{table}[ht]
    \centering
    \resizebox{\textwidth}{!}{%
    \begin{tabular}{ccccccccc}
        \toprule
        \textbf{Designation} & \textbf{LCDB Period [hours]} & \textbf{Best Period [hours]} & \textbf{Manual Period [hours]} & \textbf{Reliability Flag} & \textbf{Total N obs} & \textbf{Amplitude} & \textbf{Diameter [km]} \\
        \midrule
        588 & 7.306 & 7.306 & 4.81 & 1 & 205 & 0.152 & 130 \\
        624 & 6.924 & 6.924 & 6.92 & 3 & 306 & 0.315 & 225 \\
        \bottomrule
    \end{tabular}%
    }
    \caption{Example of the final summary table for Jupiter Trojans' rotational period analysis. This table presents two example rows from the final dataset, showing key information for each Trojan. The columns include: Designation, LCDB Period [hours], Best Period [hours], Manual Period [hours], Reliability Flag, Total N obs, Amplitude, and Diameter [km].}
    \label{tab:demo_table}
\end{table}

\section{Results and Discussion}
\label{results}

As of February 2024, about 240 reliable rotation periods of Jupiter Trojans have been reported with a completeness of about 98\% down to a diameter of 50 km \citep{Warner2009Icarus}. The most important outcome of this analysis is the identification of new rotation periods for 80 of the 216 Trojans in our sample, representing a 33\% increase in the sample size of Trojans with known rotation periods. The range of reliable rotation periods in our sample spans from 4.596 hours to 447.8 hours, as shown in Figure~\ref{fig:15033} and Figure~\ref{fig:11351}. These results are consistent with previous studies, particularly those indicating that the spin barrier for Jupiter Trojans falls between 4 and 4.8 hours \citep{2024SSRv..220...17M}. Notably, our analysis uncovered three new Trojans with rotation periods in this spin barrier range, doubling the number of Trojans known to occupy this critical period range \citep{Warner2009Icarus}.

The analysis of the rotation periods for our sample of 216 Trojans reveals a clear clustering of objects around the 4-4.8 hour spin barrier, supporting the hypothesis that Jupiter Trojans possess lower bulk densities than main-belt asteroids. 

The critical rotation period for an object to remain gravitationally bound is linked to its bulk density. For cohesionless, rubble-pile bodies, the critical density can be estimated using Equation \ref{eq:critical_density} (derived from Equation (6) in \cite{PRAVEC200012}), where $\rho$ is the bulk density (kg/m$^3$) and $P_{\text{min}}$ is the minimum observed rotation period (in seconds).

\begin{equation}
\rho = \left( \frac{3.3 \, \text{hr}}{P_{\text{min}}} \right)^2,
\label{eq:critical_density}
\end{equation}

We applied Equation~\ref{eq:critical_density} using the minimum detected rotation period of 4.596 hours for the fastest-rotating Jupiter Trojan (15033 or 1998 VY29). This calculation yields a mean density of \(\rho = 0.52 \text{ g/cm}^3\). 
The uncertainty of our estimate of the mean density given by Equation \ref{eq:critical_density} is dominated by the uncertainty of estimating the minimum period (i.e. the “spin barrier”). The latter uncertainty is driven by the sample size rather than the period uncertainty of individual objects, and we estimate it to be about 5-10\% (that is, the estimated range is 4.0 to 4.8 hours). Therefore, the uncertainty of the mean density is about 10-20\%, which makes the implied difference compared to main-belt asteroids highly significant.
It is important to note that this derived mean density using Equation \ref{eq:critical_density} is lower than the measured densities of individual Trojans, including Eurybates ($1.1 \, \mathrm{g/cm^3}$; \citep{2021PSJ.....2..170B}), Patroclus ($0.8 \, \mathrm{g/cm^3}$; \citep{2006Natur.439..565M}), and Hektor ($1.0 \, \mathrm{g/cm^3}$; \citep{2014ApJ...783L..37M}). This discrepancy suggests that the spin barrier alone may not provide a complete picture of bulk density.
Such a low density suggests that Trojans are composed of a mix of ice and volatile materials, much like cometary nuclei, rather than primarily rocky compositions typical of main-belt asteroids. This stark difference in density further reinforces the notion that Jupiter Trojans are a distinct population with unique physical characteristics, as seen in the distribution of rotation periods, which includes a concentration near the spin barrier and a noticeable lack of objects with shorter periods.

The overall distribution of rotation periods, as shown in Figure~\ref{fig:summary}, indicates that most Trojans in our sample have periods between 6 and 50 hours, with the fastest rotating Trojans clustering near the 4-4.8 hour spin barrier. This clustering supports the theory that many Trojans are rubble-pile structures, whose cohesion relies on gravity rather than material strength. The lack of Trojans with rotation periods shorter than 4 hours suggests that these objects cannot rotate more quickly without disrupting their structure. Interestingly, the data also reveal a number of Trojans with rotation periods exceeding 100 hours, hinting at the presence of binary or contact binary systems, a hypothesis supported by previous studies of slow rotators (\cite{Sonnett2015}); \cite{Ryan2017}; \cite{2021ApJS..254....7K}.

In addition to the clustering around the spin barrier, our analysis, as illustrated in Figure~\ref{fig:spin_barrier}, shows that objects near this boundary are likely rubble-pile bodies, held together by gravitational forces rather than cohesive materials. This is consistent with findings in prior works, such as \citep{2017A&A...599A..44S}, where the spin barrier was highlighted as a critical feature of the rotational distribution for Jupiter Trojans. Our data confirm this characteristic, extending previous findings by demonstrating the absence of Trojans with periods shorter than 4 hours, reinforcing the notion that Jupiter Trojans are more loosely bound and prone to disruption at higher rotational speeds. Moreover, these results echo the work of \citep{Carbognani_2017}, which discussed similar spin barrier dynamics for main-belt asteroids, although Trojans exhibit longer periods due to their lower densities.

The cumulative distribution of rotation periods, presented in Figure~\ref{fig:spin_barrier_all}, further emphasizes the distinction between Jupiter Trojans and main-belt asteroids. Trojans show a predominance of longer rotation periods. This overabundance of slow rotators suggests a bimodal distribution in rotation periods, with a primary peak near the spin barrier and a secondary peak at significantly longer periods. Such a distribution aligns with earlier studies, including \citep{2017A&A...599A..44S}, which also observed an excess of slow rotators in the Trojan population, potentially indicating a large fraction of binary systems.

\section{Conclusion}
\label{conclusion}

In this paper, we have presented new rotation periods for 80 Jupiter Trojans, increasing the known sample by 33\%. Our analysis revealed a wide range of rotation periods, from 4.6 hours to 447.8 hours, with significant clustering around the 4 to 4.8-hour range, confirming the presence of a spin barrier in this population. This spin barrier, longer than that observed for main-belt asteroids, suggests that Jupiter Trojans possess lower bulk densities, likely due to their higher proportion of ice and volatiles. Our study also identified three new Trojans with reliable periods within the spin barrier range (shown in Table \ref{tab:best_periods}), doubling the number of known objects with periods in this critical range. Using these new rotation period estimates, we derived a mean density of \( \rho = 0.52 \text{ g/cm}^3 \) with a 10-20\% uncertainty, further supporting the hypothesis that many Trojans are cohesionless rubble-pile bodies. This level of uncertainty reflects the variation in the estimated spin barrier period (4.0–4.8 hours). These findings contribute to the broader understanding of the physical characteristics and collisional history of Trojans and highlight their distinct properties compared to other small bodies in the Solar System.

\begin{table}[h!]
\centering
\begin{tabular}{cc}
\textbf{Designation} & \textbf{Best Period [hours]} \\
39264 & 4.346 \\
15033 & 4.596 \\
60401 & 4.732 \\
\end{tabular}
\caption{New rotation period measurements for three Jupiter Trojans with rotation periods within the spin barrier range (4-4.8 hours).}
\label{tab:best_periods}
\end{table}

Looking ahead, the Rubin Observatory Legacy Survey of Space and Time (LSST, \cite{2019ApJ...873..111I}) is poised to revolutionize the study of Jupiter Trojans by providing time-resolved photometry for an unprecedented number of objects. LSST's wide field of view, deep imaging capabilities, and repeated observations will enable the discovery and characterization of several hundred thousand Trojans down to a size limit of about 1 km. With such a large sample, future studies will have the statistical power to refine our understanding of the spin barrier, investigate how rotation periods vary with size, color, and dynamical class, and examine whether there are distinct sub-populations within the Trojan swarms.

Moreover, LSST's long baseline of observations will allow for more precise measurements of lightcurves, leading to better constraints on shape, density, and internal structure. This will be crucial for exploring the collisional and dynamical evolution of the Trojans, shedding light on their formation processes and their relationship to other small body populations, such as the Kuiper Belt objects and main-belt asteroids. Additionally, the ability to detect faint, small Trojans will help to complete the inventory of this population, enabling a more comprehensive understanding of their role in the early Solar System \citep{2009arXiv0912.0201L}.

In conclusion, LSST will not only build upon the foundations laid by current surveys, including ZTF, but also push the boundaries of Trojan studies by providing unparalleled data that will help answer key questions about their physical properties and origins. As more data become available, we expect to uncover new details about the formation and evolution of Jupiter Trojans, ultimately contributing to a deeper understanding of the small bodies in our Solar System and their significance in planetary science.

\section*{Acknowledgements}
The data used in this paper were obtained with the Samuel Oschin Telescope 48-inch and the 60-inch Telescope at the Palomar Observatory as part of the ZTF project. ZTF is supported by the National Science Foundation under Grants No. AST-1440341 and AST-2034437 and a collaboration including current partners Caltech, IPAC, the Weizmann Institute of Science, the Oskar Klein Center at Stockholm University, the University of Maryland, Deutsches Elektronen-Synchrotron and Humboldt University, the TANGO Consortium of Taiwan, the University of Wisconsin at Milwaukee, Trinity College Dublin, Lawrence Livermore National Laboratories, IN2P3, University of Warwick, Ruhr University Bochum, Northwestern University and former partners the University of Washington, Los Alamos National Laboratories, and Lawrence Berkeley National Laboratories. Operations are conducted by COO, IPAC, and UW.

\section{Declaration of generative AI in scientific writing}

In preparing this work, we utilized ChatGPT to refine the grammar and enhance the flow of paragraphs in this paper. Following the use of this tool, we thoroughly reviewed and edited the content as necessary, and we assume full responsibility for the final publication.

\appendix

\section{Data Availability}

This GitHub repository (https://github.com/ZhuofuLi/ZTF-Jupiter-Trojan-Rotation-Periods) provides rotation period analysis for 216 Jupiter Trojans using ZTF data. It includes a summary table with periods, amplitudes, diameters, and reliability flags. Light curve plots and data are available upon request.

\bibliographystyle{elsarticle-harv} 
\bibliography{example}






\end{document}